\documentclass[preprint2]{aastex63}
\usepackage{longtable}
\usepackage{ulem}
\usepackage{xcolor}
\usepackage{amsmath}

\newcommand{\figref}[1]{Figure~\ref{#1}}

\newcommand{\SNeIa}{SNe~Ia}

\newcommand{\C}[1]{\ensuremath{{}^{#1}{\rm C}}}
\newcommand{\Ox}[1]{\ensuremath{{}^{#1}{\rm O}}}
\newcommand{\Ne}[1]{\ensuremath{{}^{#1}{\rm Ne}}}

\newcommand{\Ni}[1]{\ensuremath{{}^{#1}{\rm Ni}}}

\newcommand{\Fe}[1]{\ensuremath{{}^{#1}{\rm Fe}}}
\newcommand{\code}[1]{\textsc{#1}}
\newcommand{\FLASH}{\code{FLASH}}
\newcommand{\CASTRO}{\code{CASTRO}}
\newcommand{\MESA}{\code{MESA}}

\newcommand{\pv}{\ensuremath{\phi}}
\newcommand{\unitspace}{\ensuremath{\,}}
\newcommand{\usp}{\unitspace}

\newcommand{\unitstyle}[1]{\ensuremath{\mathrm{#1}}}
\newcommand{\power}[2]{\ensuremath{{#1}^{#2}}}

\newcommand{\centi}{\unitstyle{c}}
\newcommand{\kilo}{\unitstyle{k}}

\newcommand{\meter}{\unitstyle{m}}

\newcommand{\second}{\unitstyle{s}}

\newcommand{\Kelvin}{\unitstyle{K}}
\newcommand{\K}{\Kelvin}  %degrees Kelvin

\newcommand{\cm}{\centi\meter}
\newcommand{\gram}{\unitstyle{g}}

\newcommand{\grampercc}{\gram\usp\power{\cm}{-3}} %mass density
\newcommand{\Msun}{\ensuremath{M_\odot}}

\newcommand{\km}{\kilo\meter}   %kilometers
\newcommand\blust{\bgroup\markoverwith{\textcolor{blue}{\rule[0.5ex]{2pt}{0.4pt}}}\ULon}

\submitjournal{ApJ}

\shorttitle{Hybrid Ia Progenitors}
\shortauthors{Augustine et al.}

\begin{document}

\title{Type Ia Supernova Explosions from Hybrid Carbon-Oxygen-Neon White
Dwarf Progenitors That Have Mixed During Cooling}

\correspondingauthor{Alan C. Calder}
\email{alan.calder@stonybrook.edu}

\author[0000-0002-9741-5987]{Carlyn N. Augustine}
\affiliation{
  Department of Physics and Astronomy \\
  The University of Alabama \\ 
  Tuscaloosa, AL, 35487-0324, USA}

\author[0000-0003-2300-5165]{Donald E. Willcox}
\affiliation{
  Center for Computational Sciences and Engineering \\
  Lawrence Berkeley National Lab \\
  Berkeley, CA 94720, USA}

\author{Jared Brooks}
\affiliation{
  Department of Physics \\ 
  University of California \\
  Santa Barbara, CA 93106, USA}

\author[0000-0002-9538-5948]{Dean M. Townsley}
\affiliation{
  Department of Physics and Astronomy \\
  The University of Alabama \\ 
  Tuscaloosa, AL, 35487-0324, USA}

\author[0000-0001-5525-089X]{Alan C. Calder}
\affiliation{
  Department of Physics and Astronomy \\
  Stony Brook University \\
  Stony Brook, NY, 11794-3800, USA}
\affiliation{
  Institute for Advanced Computational Sciences \\
  Stony Brook University \\
  Stony Brook, NY, 11794-5250, USA}

\begin{abstract}
The creation of ``hybrid" white dwarfs, made of a C-O
core within a O-Ne shell has been proposed, and studies indicate that
ignition in the C-rich central region makes these
viable progenitors for thermonuclear (type Ia) supernovae.
Recent work found that the C-O core is mixed with the surrounding O-Ne
as the white dwarf cools prior to accretion, which results in lower central
C fractions in the massive progenitor than previously assumed. To further
investigate the efficacy of hybrid white dwarfs as progenitors of
thermonuclear supernovae, we performed simulations of thermonuclear
supernovae from a new series of hybrid progenitors that include the
effects of mixing during cooling. The progenitor white dwarf model
was constructed with the one-dimensional stellar evolution code MESA and
represented a star evolved through the phase of unstable interior mixing
followed by accretion until it reached conditions for the ignition of
carbon burning. This MESA model was then mapped to a two-dimensional initial
condition for explosions simulated with \FLASH. For comparison, similar
simulations were performed for a traditional C-O progenitor
white dwarf. By comparing the yields of the explosions, we find that, as
with earlier studies, the lower C abundance in the hybrid progenitor compared
to the traditional C-O progenitor leads to a lower average yield of \Ni{56}.
Although the unmixed hybrid WD showed a similar decrement also in total iron
group yield, the mixed case does not and produces a smaller fraction of iron group elements in the form of \Ni{56}.
We attribute this to the higher central density required for ignition and the location, center or off-center, of deflagration ignition.
\end{abstract}

\keywords{Hydrodynamics, Nuclear abundances, Nucleosynthesis,
Type Ia supernovae, White dwarf stars}

%%%%%%%%%%%%%%%%%%%%%%%%%%%%%%%%%%%%%%%%%%%%%%%%%%%%%%%%%%%%%%%%%%
\section{Introduction}
\label{sec:intro}

Thermonuclear (Type Ia) supernovae (\SNeIa) are bright stellar explosions
thought to occur when approximately 1.0 \Msun\ of material composed principally
of C and O burns under degenerate conditions. This class of supernovae is
known to synthesize much of the Fe-group elements found in the galaxy, and
the light curves of these events have a special property that allows
standardization
as distance indicators for cosmological studies~\citep{phillips:absolute}.
This use resulted in the discovery of the acceleration of the expansion of
the Universe and thus the inference of Dark
Energy~\citep{riess.filippenko.ea:observational,
perlmutter.aldering.ea:measurements,leibundgut2001}, and these events
remain critical distance indicators for cosmological studies~\citep{weinbergetal2013}.
The special property of the light curve is thought to follow
from the fact that the source of luminosity, the radioactive decay
of \Ni{56} synthesized by the thermonuclear burning, is also the
principal source of opacity~\citep{Pinto2001The-type-Ia-sup}, giving
the Phillips relation between the brightness of an event and the
rate of decline of the B-band magnitude from maximum~\citep{phillips:absolute}.

Supernovae are classified observationally
by their light curves and spectra~\citep{minkowski41,bertola64,porterfilippenko87,
wheelerharkness1990conf,Fili97}, with the type Ia designation following from
the absence of H in the spectrum and the presence of a specific Si
line~\citep{filippenko:optical,hillebrandt.niemeyer:type}. These events
have been associated with C burning under degenerate conditions
for some time~\citep{hoylefowler60,arnett.truran.ea:nucleosynthesis},
but discerning the setting(s) of these events is proving difficult
and remains the subject of active research. At present there are three
widely-accepted scenarios: the {\em single degenerate} scenario,
the {\em double detonation} or {\em sub-Chandrasekhar} scenario, and
the {\em white dwarf merger} scenario.
We briefly describe these in the subsection that follows.
Also see~\citet{hillebrandt.niemeyer:type,howell2011,hillebrandtetal2013,calderetal2013,SeitenzahlTownsley2017,roepkesim2018}
for additional discussion. We note that hybrid white dwarfs are particularly 
interesting to the single degenerate picture because the greater mass of a hybrid WD might resolve a
concern.

\subsection{Proposed Progenitor Settings}\label{sec:progenitors}

The single degenerate picture posits a white dwarf gaining mass
from a companion, and the process relies on a long
period of accretion combined with either steady burning or a
series of nova explosions for a traditional
C-O WD to gain the $\sim 0.4$ \Msun\ needed for it to approach the
limiting Chandrasekhar mass~\citep{starrfieldetal2012}. As the WD approaches
the Chandrasekhar limit, conditions in the compressed core are right
to ignite the thermonuclear burning that will incinerate the star.
Within this progenitor setting, models that best reproduce observations 
are those in which the burning begins as a subsonic deflagration
during which the star expands that is then followed by supersonic 
detonation~\citep{Khokhlov1991Delayed-detonat,HoefKhok96,GameKhokOran05,
hillebrandtetal2013,calderetal2013}. We simulate thermonuclear explosion 
properties assuming this scenario and describe our methodology in
detail below.

The double detonation picture also has a white dwarf gaining mass from 
a companion. The supernova begins with 
a detonation occurring in an accreted layer of He, and that subsequently triggers 
another detonation in the underlying white dwarf \citep{woosleyweavertaam80,taam80a,taam80b,
nomoto80,nomoto82b}. Early studies indicated that this scenario could work 
for a wide range of white dwarf masses, not just the near-Chandrasekhar case 
\citep{livne90}, hence the moniker ``sub-Chandrasekhar'' also applied to this 
scenario \citep{ww94}. The mass of accreted He
is a concern because heavier
elements synthesized in the He detonation will appear in
the outer ejecta, which does not match observations~\citep{HoefKhok96,
hoeflichetal96,finkhillebrandtroepke2007,simetal2010}.  \citet{bildstenetal2007},
however, found that fairly thin He layers could flash on sub-Chandrasekhar
mass white dwarfs, encouraging
further research~\citep{simetal2012,brooksetal2015,shenetal2018,
glasneretal2018,Townsleyetal2019}.

The white dwarf merger progenitor picture has two white dwarfs coming
together and subsequently exploding~\citep{tutukovyungelson76,tutukovyungelson79,
webbink84,ibentutukov84}. This scenario provides an abundance of degenerate fuel,
which may explain some bright events~\citep{scalzo:2010,Yuan:2010}.
Early modeling efforts found that as the stars merge, the more massive white 
dwarf can ignite near its edge and fail to produce a 
supernova~\citep{saionomoto1985,saionomoto2004,Shenetal12}. 
Subsequent work allayed this concern by demonstrating
that a low accretion rate from the disrupted secondary, $ < 3 \times 
10^{-6}$ \Msun\ per year, did not heat the primary enough for ignition
\citep{kawai1987,saionomoto2004,yoonetal2007,lorenaguilaretal2009, 
pakmoretal2012b}. Contemporary
research focuses on variations on the merger idea, including inspiraling pairs,
collisions, violent mergers, and the ``core-degenerate'' model in which the merger
takes place in a common envelope~\citep{raskinetal2009,pakmoretal2011,kashi:2011,
pakmoretal2012a,Shenetal12,katzetal2016,brooksetalfast2017}.

\subsection{The Deflagration to Detonation Transition Mechanism Within
the Single Degenerate Scenario}

The approach we employ for this study is a variation of the delayed
detonation described above, the deflagration-to-detonation transition (DDT)
explosion paradigm~\citep{1986SvAL,
Khokhlov1991Delayed-detonat,NiemWoos97,Niem99,belletal2004,fishjump2015}.
In this case, the accretion of mass on the white dwarf compresses and
heats the core, igniting carbon fusion and driving a period of convection
\citep{WoosWunsKuhl04,wunschwoosley2004,Kuhletal06,nonakaetal2012}.
At some point, the fusion rate becomes fast enough due to the rising
temperature that energy production exceeds convective cooling and
the deflagration phase begins in the core~\citep{Nomo84,WoosWunsKuhl04}.

This flame is unstable, and as the deflagration propagates toward the surface
of the WD, it is subject to the Rayleigh-Taylor instability that generates
turbulence and boosts burning~\citep{taylor+50,chandra+81}.
Burning proceeds as a deflagration for about one second, and then
the flame transitions to a detonation~\citep{hoflich.khokhlov.ea:delayed}.
Our simulations assume that the transition occurs when the top of a
rising, Rayleigh-Taylor unstable plume reaches a characteristic low
density~\citep{townsley.calder.ea:flame}. 

In the DDT paradigm, the duration of the deflagration phase sets
the amount of expansion of the star prior to the bulk of burning,
which is critical to the composition of material synthesized in the 
explosion. Also, early burning during the deflagration phase is at 
high enough densities that the effects of electron capture are 
significant and similarly influence the composition of the material
synthesized in the explosion~\citep{hoeflichetal2004,
hoeflich2006,fesenetal2007,diamondetal2018}.

\subsection{A Recent Advance in Stellar Evolution: Hybrid White Dwarfs}

Modern computing resources now enable simulations with unprecedented
realism, allowing both one-dimensional simulations with a vast
amount of included physics and full three-dimensional simulations
albeit with less included physics~\citep[][and references therein]{caldertownsley2018}.
In the area of stellar evolution, recent investigations revisiting
late-time evolution of roughly 8 \Msun~stars indicate that under
the right circumstances, ``hybrid" white dwarfs having a C-O core surrounded by O-Ne
mantle may form \citep{siess2009,denissenkovetal2013}. These hybrid
white dwarfs are thought to form when mixing at the lower convective boundary quenches
C burning in an asymptotic giant branch (AGB) star, leaving unburned C
in the core. The situation is at best uncertain, however, and the
results depend on assumptions about convective
overshoot that have been questioned~\citep{chenetal2014,lecoanetetal16,lattanzioetal2017}.

These hybrid WDs are relevant to the supernova problem because 
they can become the progenitors of thermonuclear supernovae
if they are part of a binary system. If the companion star becomes
another WD, the two can merge and produce an explosion.
If, on the other hand, the hybrid WD has a main sequence 
or giant companion, it can gain mass and approach the Chandrasekhar mass, i.e.\  
the single degenerate picture~\citep{willcoxetal2016}.

In any case, the hybrid WD will experience a period of cooling,
which will reduce the stabilizing temperature gradient (the O-Ne layer is initially hotter 
than the C-O core) and allow the unstable composition gradient to drive 
thermo-compositional convection~\citep{brooksetal2017,schwabgaraud2019}.
In the case of the
hybrid WD accreting and approaching the Chandrasekhar mass, accretion will
heat the core and start C fusion, leading to a period of ``simmering" prior
to the explosion, also mixing the interior \citep{PiroBild08}. The upshot is 
that there is likely to be considerable mixing after the hybrid forms that will 
homogenize the composition \citep{denissenkovetal2015,brooksetal2017,schwabgaraud2019}.

Hybrid WDs have more mass than traditional C-O WDs, with some studies indicating the
mass can approach 1.3 \Msun\ \citep{chenetal2014}. This increased mass
minimizes one of the problems associated with the single-degenerate picture,
the need to accrete enough mass for the WD to approach the Chandrasekhar
mass~\citep{chenetal2014,denissenkovetal2015,kromeretal2015}.
Accordingly, there has been considerable interest in viability of explosions from
these progenitors.

From population synthesis, \citet{mengpods2014} found that these
%meng received may 5 accepted may 16
progenitors may substantially contribute to the population of \SNeIa\ (1-8\%) and have
relatively short delay times. They also suggested that these
may produce part of the Iax class of events. \citet{Wangetal2014}, also with population
synthesis, studied the case
%received august 28 accepted sept 26
of a hybrid progenitor accreting from a non-degenerate He star and found
birth rates indicating that up to 18\% of SNe Ia may follow from this channel
and very short delay times. \citet{Wangetal2014} also suggested that explosions
from hybrid progenitors may provide an explanation for type Iax events.
\citet{mengpods2018}, from the common-envelope-wind model developed in
\citet{mengpods2014}, propose that both Ia-CSM and Iax events
are caused by the explosion of hybrid progenitors, with Ia-CSM occurring in systems with
a massive common envelope and Iax events occurring in systems where most of the common envelope
has been lost.

Other groups have simulated explosions from hybrid progenitors.
\citet{kromeretal2015} performed pure deflagration simulations from models
with a C core. They found that their models may explain some faint events
such as SN 2008ha~\citep{foleyetal2009}.
Interpretation of these results are made challenging by their use of a progenitor in which no mixing has occurred, not even that expected due to simmering.
\citet{bravoetal2016} performed one-dimensional simulations of explosions from
a variety of progenitor models assuming both pure deflagration and the DDT
explosion mechanism.
Some of their models are similar to those of \citep{denissenkovetal2015} and
they report that many models produce less synthesized
\Ni{56}, indicating dimmer events. They also note that some of their
models may explain Iax events.
\citet{willcoxetal2016} simulated explosions from the progenitors of
\citet{denissenkovetal2015} and found lower
\Ni{56} yields on average and a trend of lower ejecta kinetic energy
when compared to explosions from traditional C-O progenitors.
We will compare to the work of \citet{willcoxetal2016} extensively below.
They did not allow for the mixing of the WD during cooling, before ignition of carbon burning.
Here we utilize progenitors in which mixing during cooling has been modeled appropriately \citep{brooksetal2017}.

Discerning the role played by hybrid progenitors in the global gamut of
thermonuclear supernovae is the goal of the study we present in this paper. Our simulations
provide the yields of \Ni{56} in the supernova explosions. Because
the radioactive decay of \Ni{56} powers the light curve of an event, 
the yield of \Ni{56} serves as a proxy for brightness of the event so
our estimates allows us to discern trends in brightness. Our study thus 
can offer insight into the relative brightness of explosions from these
progenitors as well as the overall scatter in the brightness
of observed events.

Our simulation methodology is described in the next section and
our results include a comparison of nucleosynthetic explosion yields between these
new mixed hybrid C-O-Ne progenitors and traditional C-O progenitors
in the section following the methodology. Our results show systematic
differences in the yield of \Ni{56} between explosions from the two
types of progenitors, which we quantify and discuss.

\section{Methodology}
\label{sec:method}

The methodology of our study follows the approach of \citet{willcoxetal2016}.
We performed a suite of two-dimensional simulations of supernova explosions
from hybrid progenitors and compared the results to a suite of simulations
of explosions from traditional C-O progenitor models performed with the
same simulation code and from similar initial conditions. The simulations assumed
the deflagration to detonation transition explosion paradigm, and the
transition densities were the same in both suites. We briefly review our methodology
here, and refer the reader to previous work for additional details.
In particular, we use the same simulation instrument, a modified version
of the \FLASH\ code, as \citet{willcoxetal2016} and we refer the reader there
for a description of the process and treatment of burning of C--O--Ne fuel and how it differs from C--O fuel.

\subsection{Simulation Instrument}

The simulations of supernova explosions presented here were performed
with a customized version of the \FLASH\ code, originally developed
at the University of Chicago.
\FLASH\ is a parallel, adaptive mesh, multi-physics simulation code
developed first for nuclear astrophysics applications and subsequently
for high-energy-density applications~\citep{Fryxetal00,calder.curtis.ea:high-performance,
calder.fryxell.ea:on,flash_pragmatic,flash_evolution}.
\FLASH\ has been applied to a variety of astrophysical problems by a host
of researchers, and the version we apply differs from other versions
principally in the modules describing thermonuclear burning via a
flame capturing model.

The need for a model flame in simulating thermonuclear supernovae follows
from the scales of the problem. At high densities, the width of a laminar
nuclear flame is $< 1\ensuremath{\;}{\ensuremath{\mathrm{cm}}}$ while the
radius of the white dwarf $\sim 10^9\ensuremath{\;}{\ensuremath{\mathrm{cm}}}$.
Even with adaptive mesh refinement, whole-star simulations cannot
simultaneously resolve the nuclear flame, so simulations require a model to
describe the burning on unresolved scales.  The model we apply is a flame
capturing scheme and thermally-activated burning module to describe
thermonuclear burning during both the deflagration and detonation phases,
as well as routines to describe the evolution of the dynamic ash.
This description of the burning was developed during the course of research in
thermonuclear supernovae and has been presented in a series of
papers \citep[See][and references therein]{townetal2016,chandlery}. For completeness,
we briefly review the flame capturing scheme here.

For the deflagration phase, the flame capturing scheme propagates an
artificially broadened flame with an advection-diffusion-reaction (ADR)
scheme~\citep{Khok95,VladWeirRyzh06} via evolution of
a reaction progress variable to describe the consumption of C
and additional variables to describe the evolution of intermediate-mass
elements into the statistical quasi-equilibrium of the
Si-group~\citep{ifk1981,khok1981,khok1983} and then into iron-group elements
(IGEs) in full nuclear statistical equilibrium (NSE).

The reaction progress variable is $\phi$ and it varies from $\phi=0$ for
unburned fuel to $\phi=1$ for burned ash. $\phi$ is evolved
via an advection-diffusion-reaction equation,
\begin{equation}
  \label{eq:ard}
  \partial_t \pv + \vec{u}\cdot\nabla \pv = \kappa \nabla^2 \pv +
\frac{1}{\tau} R\left(\phi\right) ,
\end{equation}
where $\vec{u}$ is the velocity of the fluid, $\kappa$ is the
diffusion coefficient, $\tau$ is the reaction timescale, and $R(\phi)$ is
a non-dimensional function describing the reaction. The parameters
$\kappa$, $\tau$, and $R(\phi)$ are tuned to propagate the reaction
front at a prescribed speed.  Our model uses the ``sharpened KPP''
 described by~\cite{VladWeirRyzh06},
with $R\propto(\phi-\epsilon)(1-\phi+\epsilon)$, where
$\epsilon \simeq 10^{-3}$.  This scheme has been shown to be
acoustically quiet, stable, and to give a unique flame speed
\citep{townsley.calder.ea:flame}. The input flame speeds come from
tabulated results obtained by direct numerical simulations of
thermonuclear burning.
The flame speeds are obtained via linear interpolation within a 
three-dimensional table that combines the results of \citet{timmes92} 
and \citet{Chametal08}, and the dimensions of the table are \C{12} 
mass fraction, \Ne{22} mass fraction and log of density. While the
WD contains additional species, we have shown that treating the
abundance of \Ne{22} as a proxy for neutron-rich elements captures the
speed-up of laminar flames due to neutronization and thus reasonably 
produces variations in flame speed due to composition~\citep{jacketal2010}.
Using these tables, which were made for mixtures without $^{20}$Ne,
effectively assumes that the substitution of $^{20}$Ne for $^{16}$O will not
change the flame speeds significantly.  We consider this a reasonable
approximation given that, in the interior of our star, $^{16}$O is still much
more abundant than $^{20}$Ne, and that the flame propagation is largely
dominated by turbulent effects.
These laminar flame speeds are boosted to account for the speed-up 
of burning due to unresolved buoyancy and background turbulence~\citep{Khok95,
gamezo.khokhlov.ea:thermonuclear,townsley.calder.ea:flame,jacketal2014}.

The two-dimensional models in this study do not utilize a sub-grid-scale
model for the turbulence-flame
interaction~\citep[See][for examples]{Schmetal06a,Schmetal06b,jacketal2014}.
Those sub-grid-scale models only apply in three-dimensional
simulations because two-dimensional hydrodynamics cannot
correctly describe turbulence. The simulations we present
use the minimal enhancement based on the Rayleigh-Taylor
strength introduced by~\citet{townsley.calder.ea:flame}. The
assumption here is that the burning self-regulates on resolved
scales so that results are insensitive to the detailed treatment
of the interaction with turbulence, and previous experience
indicates this assumption is reasonable for comparisons
like the one presented in this work~\citep{townsley.calder.ea:flame,
willcoxetal2016}.

The ADR scheme describes the consumption of C and the subsequent stages
are described by separate progress variables and
separate relaxation times derived from full nuclear network calculations
\cite{Caldetal07,townetal2016}. In both the quasi-equilibrium and full equilibrium,
the creation of light elements by photodisintegration balances the creation of heavy
elements by fusion, maintaining the equilibrium. The relative balance depends on
thermodynamic conditions, e.g.\ density and temperature, and hydrodynamic motion
during the explosion changes the thermodynamic conditions and thus the
balance.
Electron capture also influences the evolution in several ways by
neutronizing the material, which produces more neutron rich iron-group
material at the expense of \Ni{56}. Neutronization also shifts the binding
energy of the material and the Fermi energy of electrons, respectively
changing the temperature (due to released energy) and the pressure.
Finally, individual electron capture reactions emit neutrinos that
escape and remove energy from the system.
Like the input flame speed, the burning model includes tabulated rates for these effects
from detailed NSE calculations ~\citep{SeitTownetal09}. Accordingly, the burning
model is able to describe dynamic evolution of the ash in addition to the
stages of C-O-Ne burning.

The burning model also describes the detonation phase with progress variables.
In this case, the model evolves thermally-activated burning with the actual
temperature-dependent rate for C consumption, which allows a propagating shock to
trigger burning, i.e.\ to propagate a detonation front. The propagating detonation
is able to describe the same stages of C burning as the deflagration case, including
the relaxation into NSE~\cite[and references therein]{townetal2016}. Finally,
we again note that the burning model was adapted for the case of burning in
hybrid white dwarfs.
Parameter studies of detonations in C-O-Ne material and details of how the burning 
model is used to capture C-O-Ne burning may be found in \citet{willcoxetal2016}.

\subsection{One-dimensional Hybrid Model}

The hybrid white dwarf model that served as the initial condition
for the simulations of supernova explosions presented in this work
was constructed with the one-dimensional stellar evolution code
MESA~\citep{mesa1,mesa2,mesa3,mesa3e}. The evolution of the model 
included larger nuclear reaction networks than previous studies and 
thus resulted in a electron-to-baryon gradient that became unstable 
to mixing as the interior cooled, leading to a lower central C 
fraction than previous hybrid models~\citep{brooksetal2017}. We 
selected the 1.09 \Msun\ WD model 
with the 0.4 \Msun\ C/O core from \citep{brooksetal2017}.
In order to grow this WD towards the Chandrasekhar mass, it was inserted into 
a binary with a 1.4 \Msun\ He star donor having a 3 hour orbital period, similar to 
the simulations in \citet{brooksetal2016},
and the system was
allowed to evolve.  The WD grows, experiences central carbon ignition and 
simmering, and then is stopped when its central temperature reaches 
$\log T_{\rm c}(\Kelvin) = 8.9$, 
at which point the 
temperature is just about high enough to ignite the deflagration that is 
the first stage of the explosion. The central \C{12} mass fraction 
of the model is 0.1419.

\begin{figure}
\includegraphics[width=\columnwidth]{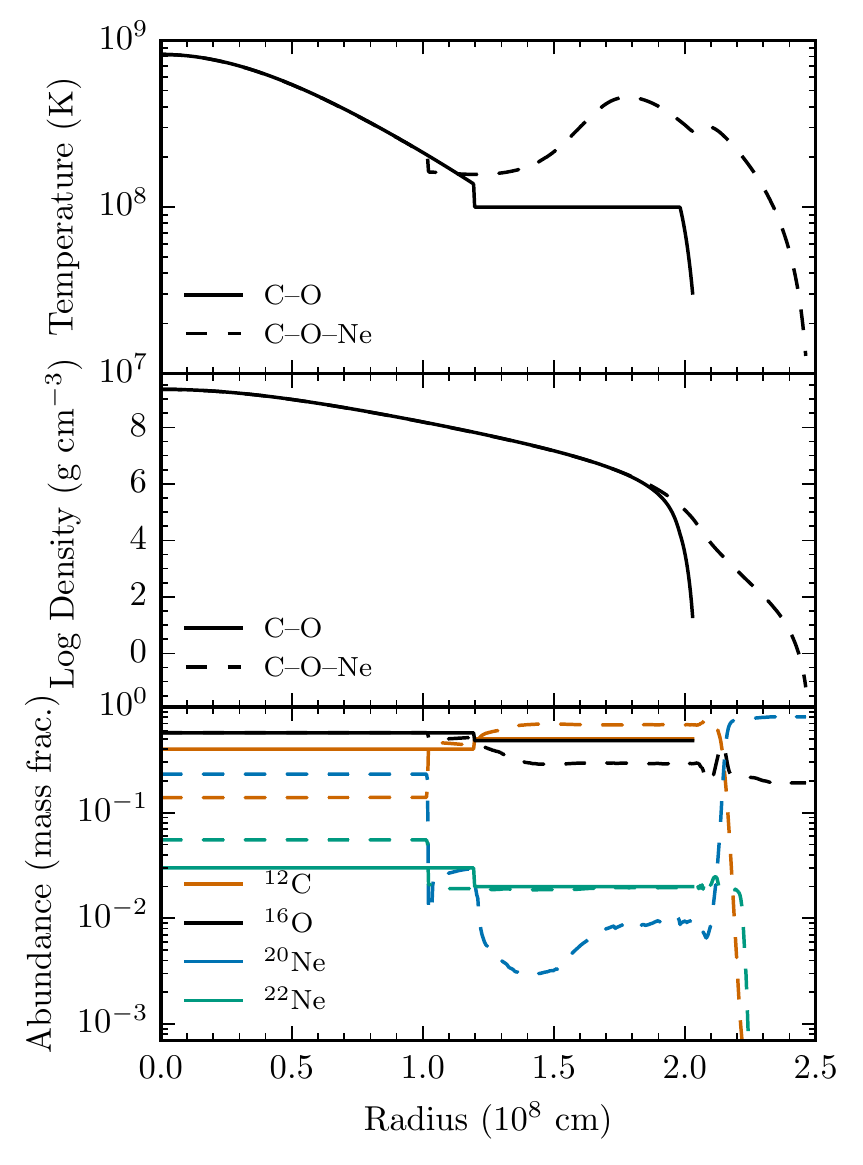}
\caption{\label{fig:init_conds}
Radial temperature, density, and  composition profiles of the one-dimensional hybrid
progenitor WD. The central temperature and density were the same as
the traditional C-O progenitor model.
The peak temperature of the hybrid model is at the center of the star,
so both the hybrid and C-O models could share the same central ignition initial conditions in our simulations.
}
\end{figure}

Following the approach of \citet{willcoxetal2016}, we constructed a
corresponding ``classic'' C-O model to allow comparison of explosion
results between these hybrid models and those of previous studies
\citep{Krueger2010On-Variations-o,Kruegeretal2012}. The C-O model was 
constructed to have conditions as similar a possible to the hybrid
model (e.g.\  the two shared the same central temperature and density).  
The central \C{12} mass fraction of this model is 0.4.
Figure~\ref{fig:init_conds} shows the density and temperature
profiles of the two initial one-dimensional models.
While the C-O-Ne progenitor is the result of an evolutionary calculation,
the C-O progenitor is constructed by integrating the equation of hydrostatic equilibrium
using the Helmholtz stellar equation of state of \citet{timmes.swesty:accuracy} starting
from the central density and temperature of the C-O-Ne WD model.
This parameterized construction is reflected in the simplified
outer thermal structure for the C-O WD model appearing in Figure~\ref{fig:init_conds}.
We centrally ignite our explosion simulations for both C-O and hybrid
models to coincide with the peak temperature at the center of the WDs,
unlike the hybrid models of~\citet{willcoxetal2016}.

\subsection{Two-dimensional Initial Conditions From One-dimensional Models}

The initial progenitor models for the two-dimensional simulations were
created from the one-dimensional \MESA\ models by mapping the one-dimensional
models onto the two-dimensional domain while
preserving hydrostatic equilibrium.
As described above, the flame capturing scheme with progress
variables describes the evolution as the material burns. The \MESA\
models, however, relied on a detailed reaction network that included
many species. Accordingly, the process of creating the two-dimensional
models for \FLASH\ required aggregating some of the species abundances.
As with much of this study, we applied the
techniques of~\citet{willcoxetal2016}.

The process of mapping the \MESA\ models began by converting
the \MESA\ model to a uniform grid of $4$~km resolution.
To do this, \MESA\ quantities were mass-weighted and averaged in zones with spacing less than $4$~km
and interpolated in zones with spacing greater than $4$~km.
At this point,
abundances of nuclides from the \MESA\ models were aggregated
into the abundances tracked by the flame capturing scheme in
\FLASH.
The most abundant isotopes in the model were \C{12}, \Ox{16},
and \Ne{20}, which are symmetric (the number of neutrons
equals the number of protons). Similarly to~\citet{willcoxetal2016},
other neutron-rich isotopes in the initial model were combined into
\Ne{22}, which serves as a proxy for metallicity. The aggregation
accounts for the $Y_e$ of the full set of nuclides, and \Ne{20}
and \Ox{16} were constrained to be in the same ratio in both
sets of abundances. \figref{fig:init_conds} shows the initial
profile of the one-dimensional models.

The original \MESA\ model was in equilibrium, but to ensure
the uniformly-gridded model was in hydrostatic equilibrium we
constructed the appropriate pressure profile. This was done by
integrating for the pressure in each zone from the central point up,
accounting for the local acceleration of gravity, temperature,
composition, and the mass below and enclosed by the zone. As
with \citet{willcoxetal2016}, we used the EOS routine from
\CASTRO\ ~\citep{timmes.swesty:accuracy,castro1}.
This procedure produced a structure that was stable in \FLASH, with 
fluctuations in central density less than $3\%$, for at least $5$ 
seconds with no energy deposition.

\subsection{DDT Process and Suites of Explosions}

The simulations performed for this study consisted of a suite of
30 two-dimensional simulations of thermonuclear supernova explosions 
from hybrid C-O-Ne progenitors in the DDT explosion paradigm. These
were compared to a suite of supernova simulations from traditional C-O 
progenitors. The C-O progenitors are parameterized and 
include the effects of convective ``simmering" in the core as the WD 
approaches the Chandrasekhar mass~\citep{Chametal08,PiroBild08,jacketal2010}.
Using parameterized C-O models allowed us to choose conditions, e.g.\
central density, to control differences between the hybrid and traditional 
models and thus assess the effect of the hybrid structure.
The central density at ignition depends on both the carbon abundance 
and the binary history, mainly cooling time before accretion 
\citep{Kruegeretal2012}, and so this comparison does not correspond 
to a comparable binary scenario, but is intended as a more 
straightforward comparison.

The C-O and C-O-Ne models have the same central temperature
and density, $8.2\times 10^8$~K and $2.2\times 10^9$~g~cm$^{-3}$ respectively.
The C-O model's composition consisted of \C{12}, \Ox{16}, and \Ne{22} in mass fractions 0.4, 0.57, and 0.02 respectively in the convective core and 0.5, 0.48, and 0.02 outside the convective core.
The C-O-Ne model used in \FLASH\ had composition consisting of \C{12}, \Ox{16}, \Ne{20}, and \Ne{22}.
The mass fractions in the central convection zone were 0.14, 0.68, 0.23, and 0.05 respectively, and vary smoothly outside the convection zone as shown in Figure~\ref{fig:init_conds}.

In both suites, the simulation
begins with a progenitor model mapped to the two-dimensional
\FLASH\ grid. The burning is initiated with a ``match head,'' a region
in the white dwarf's center that is fully burned to nuclear statistical
equilibrium (NSE).
This initially burned region ignites a deflagration, a subsonic
flame, and because the match head was perturbed it is unstable to
the Rayleigh-Taylor instability with the result that buoyant plumes
rise. The star is partially consumed during this deflagration phase,
and the star responds by expanding.
When a plume reaches a specified density,
a detonation is initiated, and the simulation continues until
the expanding star reaches low densities at which point burning
effectively ends. This section provide details
of the implementation of this method.

The simulations were performed in two-dimensional $r$-$z$
cylindrical coordinates, extending radially from 0 to
$65,536$~\km\ and along the axis of symmetry from $-65,536$~\km\ to
$65,536$~\km. The maximum refinement level of the adaptive mesh
corresponded to $4$~\km\ resolution, which previous study has shown
is a good balance between efficiency and
accuracy~\citep{townsley.calder.ea:flame,townetal2009}.
This resolution and geometry was used in previous studies
allowing direct comparison to previous results~\citep{Kruegeretal2012}.

The ignition of the deflagration via a match head followed the
initialization described in~\citet{Kruegeretal2012}, which
followed the method of~\citet{townetal2009}.
In both the hybrid and traditional cases, the match head had a
nominal radius of $150$~km before
a different randomly-seeded perturbation was applied to the
match head for each of the 30 simulations in both suites,
The perturbation to the sphere's surface is a set of
spherical harmonic functions with randomly chosen amplitudes, and
each set of perturbations is referred to as a ``realization."
Both suites use the same 30 realizations of the ignition geometry.
These perturbations have been shown to reproduce
the scatter in \Ni{56}
yield from \SNeIa\ \citep{townetal2009}.
We note that the ignition points of the hybrid models of 
\citet{willcoxetal2016} substantially differed from both the
(mixed) hybrid models presented here and  traditional models. Those progenitors
had the highest temperatures and hence were ignited at a radius of
about 300 km.

The transition from deflagration to detonation again follows
the previous studies, which assumed the transition location
is parameterized by the fuel density $\rho_{\mathrm{DDT}}$.
When a rising plume reaches the threshold density, in this case $\rho_{\mathrm{DDT}} = 10^{7.2}~\grampercc$, a $12$~\km\ radius region of fuel, is fully burned $32$~\km\ radially outwards from the flame.
This
instantaneous burning in the region of this size provides conditions
to generate the shock and support the detonation at the chosen
threshold density.  Multiple DDT points may arise, but they are
constrained to be at least $200$~\km\ apart. The choice of DDT
in the suite has been shown to be high enough to ensure
the robust ignition of a detonation shock.

Once the detonation starts, the remaining fuel at densities
high enough for the detonation to propagate is quickly consumed.
The simulations were run to 4.0 s, by which time burning
has effectively ceased.

\section{Results}

We frame the presentation of the results of the suites of simulations
principally in terms of the yield of \Ni{56}, the energy source of
the light curve of an event. \Ni{56} thus serves as a proxy
for the brightness of an explosion and comparison of the yields
is equivalent to comparing the brightness of events. The
yields were estimated from $Y_e$ and the NSE progress variable, by 
assuming the composition upon NSE freeze-out is \Ni{56} plus equal
parts \Fe{54} and \Ni{58} \citep{townetal2009,Meaketal09}.
This assumption allows the fraction of IGEs in the form of \Ni{56} to be 
estimated from the $Y_e$ tracked by the burning model. 
This process has been shown to provide estimated \Ni{56} yields consistent with
the results of detailed network and NSE calculations~\citep{townetal2016,caldertownsley2018}.

\begin{figure}
\includegraphics[width=\columnwidth]{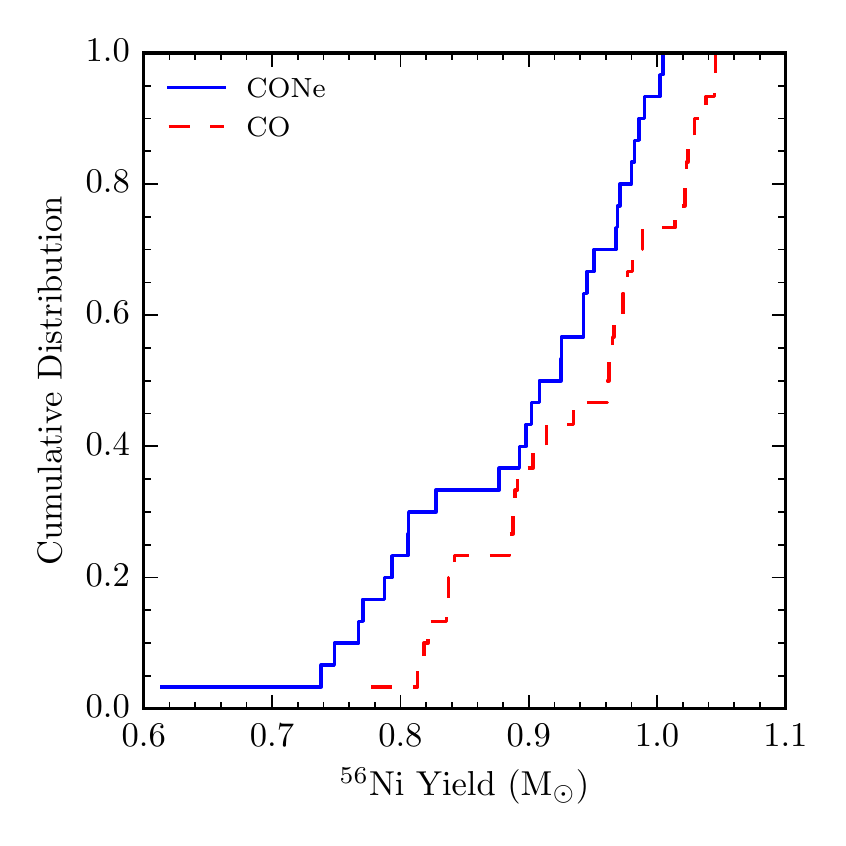}
\caption{\label{fig:cumdist}
The cumulative distribution of the final \Ni{56} yield for explosions from
C-O (red) and hybrid C-O-Ne (blue) models.
The curve for the hybrid models is shifted
to the left, indicating that explosions from hybrid progenitors produce
less \Ni{56} than explosions from traditional C-O models with the same central density.
}
\end{figure}
The cumulative distribution of the \Ni{56} yield for explosions from
C-O and hybrid C-O-Ne models is presented in Figure \ref{fig:cumdist}.
Average and standard deviations of the sample of yields and kinetic energies are given in Table~\ref{tab:ave_yields}.
The figure shows that the hybrid models consistently have a higher
cumulative fraction at a given mass of \Ni{56}, which indicates
that the yields of \Ni{56} are consistently lower in explosions
from hybrid progenitors of the same central density.
This contrast is in a similar direction but not as large as the difference 
between C-O and C-O-Ne progenitors seen in \citet{willcoxetal2016}.

\begin{table}
\caption{\label{tab:ave_yields}Average Yields and Kinetic Energies}
\begin{tabular}{lccc}
\hline
%Progenitor & $^{56}$Ni & IGE & Kinetic Energy \\
Progenitor & \Ni{56} & IGE & Kinetic Energy \\
 & ($M_\odot$) & ($M_\odot$) & ($10^{51}$~erg)\\
\hline
C--O     & $0.94\pm0.08$ & $1.14\pm0.08$ & $1.35\pm0.06$ \\
C--O--Ne & $0.89\pm0.10$ & $1.15\pm0.11$ & $1.21\pm0.07$ \\
\hline
\end{tabular}
\end{table}

\begin{figure}
\includegraphics[width=\columnwidth]{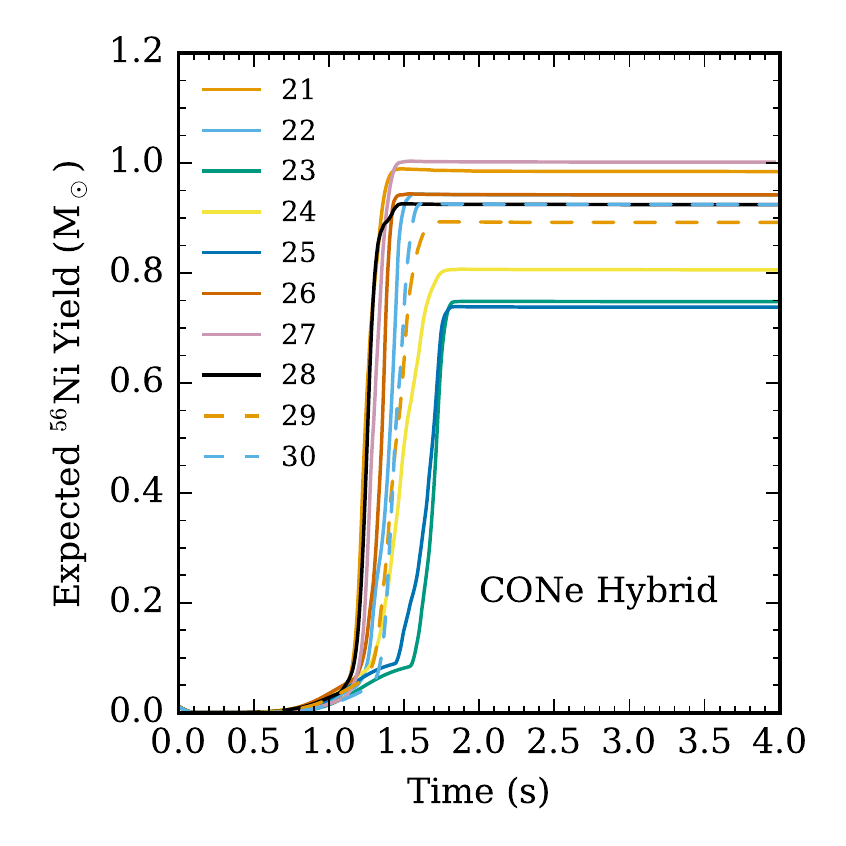}
\includegraphics[width=\columnwidth]{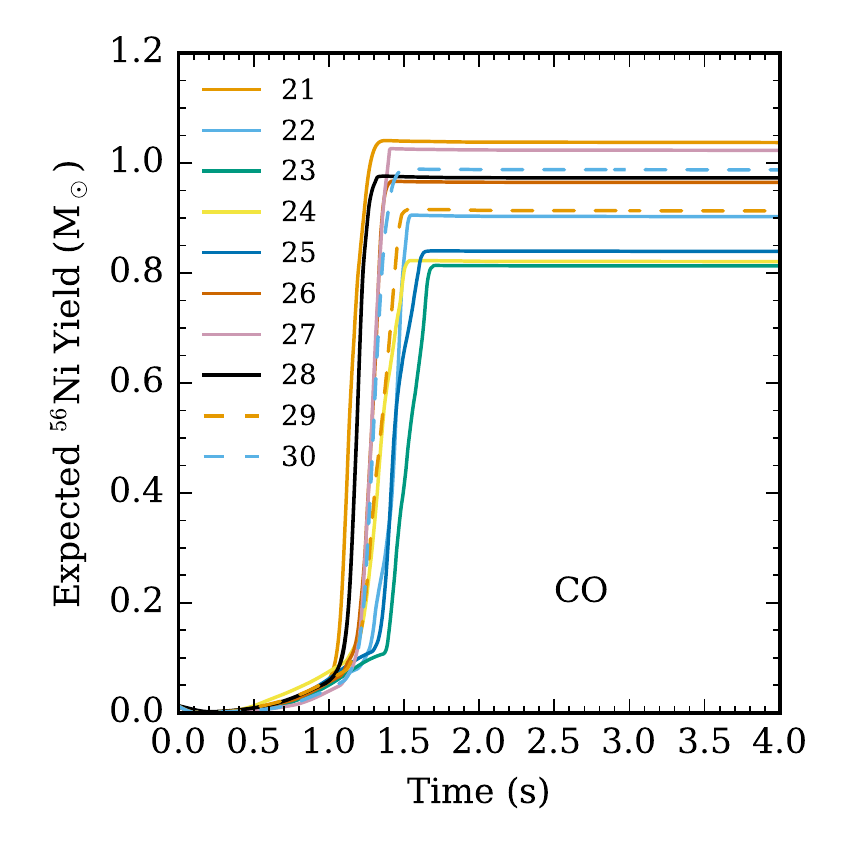}
	\caption{\label{fig:nithybrid} \label{fig:nitco}
Evolution of expected \Ni{56} yield in time for ten hybrid C-O-Ne WD (top panel) and ten C-O WD (bottom panel) explosion simulations.
Shown are realizations 21-30.
}
\end{figure}

The production of \Ni{56}, after NSE freeze-out, as a function of simulation time
for 10 simulations of explosions from each progenitor is shown in Figure \ref{fig:nithybrid}.
The sharp increase in the yield occurring after approximately 1.0 second indicates
the transition to the detonation phase with its significantly faster burning. The curves
indicate that on average, the C-O-Ne models reach the DDT later then the hybrid model, 
implying more expansion of the WD and lower density burning in the detonation phase.

\begin{figure}
\includegraphics[width=\columnwidth]{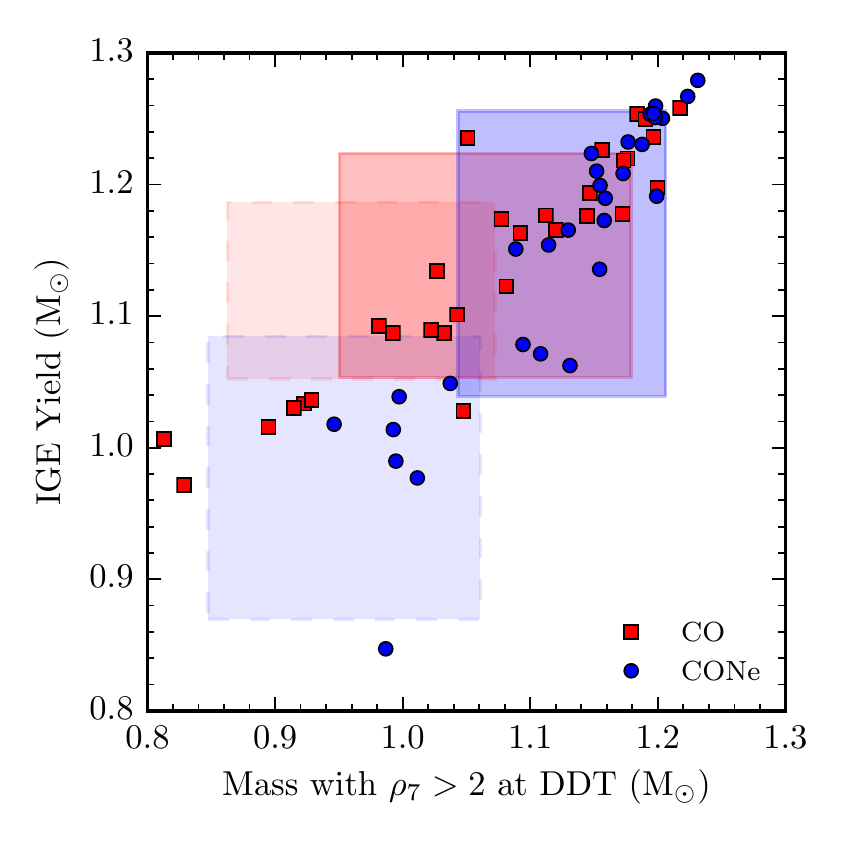}
\caption{\label{fig:masshighdens}
Final IGE yield
vs.\ mass above $2\times 10^7~\grampercc$
at the time of the first DDT in a simulation.  Individual
realizations are shown (C-O-Ne
progenitors in blue, C-O in red) along with rectangles of length
$\pm 1 \sigma$ along each axis and centered at their average values
for C-O and C-O-Ne.  Similar rectangles found in \citet{willcoxetal2016} for a progenitor without
mixing during cooling are shown in a lighter shade with dashed borders.
}
\end{figure}

Figure \ref{fig:masshighdens} compares the final yield of IGEs
for the C-O and C-O-Ne simulation suites with the amount of expansion of the
WD during the deflagration phase. The amount of expansion is
characterized by the mass above $2 \times 10^7~\grampercc$ at the
time of the first DDT occurrence, with more high-density mass
indicating less expansion during deflagration \citep{townetal2009}.
The averages and standard deviations, $\sigma$, of both
the C-O and C-O-Ne suites along both axes are indicated by the shaded
regions with $\pm1\sigma$ widths.
Lighter shaded regions indicate the averages and standard deviations found by \citet{willcoxetal2016} for hybrid C-O-Ne progenitors that do not mix during cooling and for C-O progenitors with the same central density as them.

As expected from previous studies \citep[e.g.][]{townetal2009}, the trend for both C-O and C-O-Ne
models is that less expansion during the deflagration phase results in
greater IGE yields.
This is because less expansion allows the detonation to consume more high density fuel, which is burned more completely \citep[See discussion in][and references therein]{SeitenzahlTownsley2017}.
At moderate expansion, where the mass at high density is between around 1.0 to 1.1 M$_\odot$, our C-O models tend to yield greater IGE mass.
Following
\citet{willcoxetal2016}, we interpret
the lower IGE yield in C-O-Ne models in this range as resulting
from the lower \C{12} abundance and the fact that, given
similar fuel density, the \Ne{20}-rich fuel will burn to cooler
temperatures than fuel in the C-O models. The result is
slower burning to IGE and thus a lower IGE yield.
At lower degrees of expansion, where there is around 1.2~M$_\odot$ of material at high densities, there appears to be little difference between the IGE yields of C-O and C-O-Ne progenitors.
This convergence for dense, weakly expanded, cases was not noted in \citet{willcoxetal2016} because they had very few cases with more than 1.1~M$_\odot$.
It does appear consistent with an extrapolation of their data and their two cases that did yield these higher masses at high density.

This convergence of IGE yield also appears strongly in comparing the averages over the whole set found in this study and that of \citet{willcoxetal2016}.
As seen from the shaded regions in Figure \ref{fig:masshighdens} (lighter shade is unmixed), the degree of expansion seen for the our mixed progenitors is significantly less than for the unmixed cases.
This results in more mass at high density in the mixed cases, leading the C-O-Ne case to produce a similar amount of IGE, on average, to the C-O case.

There are two major differences between the explosions computed in \citet{willcoxetal2016} and here.
First, the central density of the progenitor that mixed during cooling is about 60\% higher than that of the unmixed one, $2.2\times10^9$~g~cm$^{-3}$ compared to $1.4\times 10^9$~g~cm$^{-3}$.
This is due to the higher central density necessary to ignite carbon burning with the lower central carbon fraction resulting from mixing during cooling
(\citealt{brooksetal2017}; Section \ref{sec:method} above).
Second, while the case without mixing during cooling led to an off-center ignition of the deflagration, here the mixed case ignited at the center.
We conclude that each of these contribute in specific ways to the differences between the results of mixed and unmixed progenitors.

Since the mass at high density also increases with increasing central density for the C-O progenitor,
it appears likely that the difference in central density is important for this increase in the C-O-Ne case as well.
However, the increase in the mass at high density for the C-O-Ne case is more pronounced.
This suggests that the central ignition also plays a role.
The increase in the IGE yield that results from this higher central density is negligible for the C-O case.
This is consistent with the results of \citet{Kruegeretal2012}, who found that IGE yield is fairly independent of central density.
The C-O-Ne case, however, appears to act very differently.
The mean IGE yield found here is more than 0.15~M$_\odot$ larger than that found in \citet{willcoxetal2016}.
This increase closes the gap, making the average IGE yields from the C-O and C-O-Ne cases similar.
The steeper dependence of IGE yield on expansion is partially responsible for this large change, in combination with the larger change in how much expansion happens in the C-O-Ne case.
This supports both the central density and the location of ignition playing a role in bringing the explosions into a range where C-O and C-O-Ne cases produce similar amounts of IGE.

While the higher central density makes the IGE yields quite similar, the hybrid C-O-Ne WD still produces less \Ni{56} than a C-O WD at similar central density.
In order to understand this we looked more carefully at how the process of electron capture during the explosion proceeds in our simulations.
Again we find that both the central density of the progenitor and the location of ignition are important.

\begin{figure}
\includegraphics[width=\columnwidth]{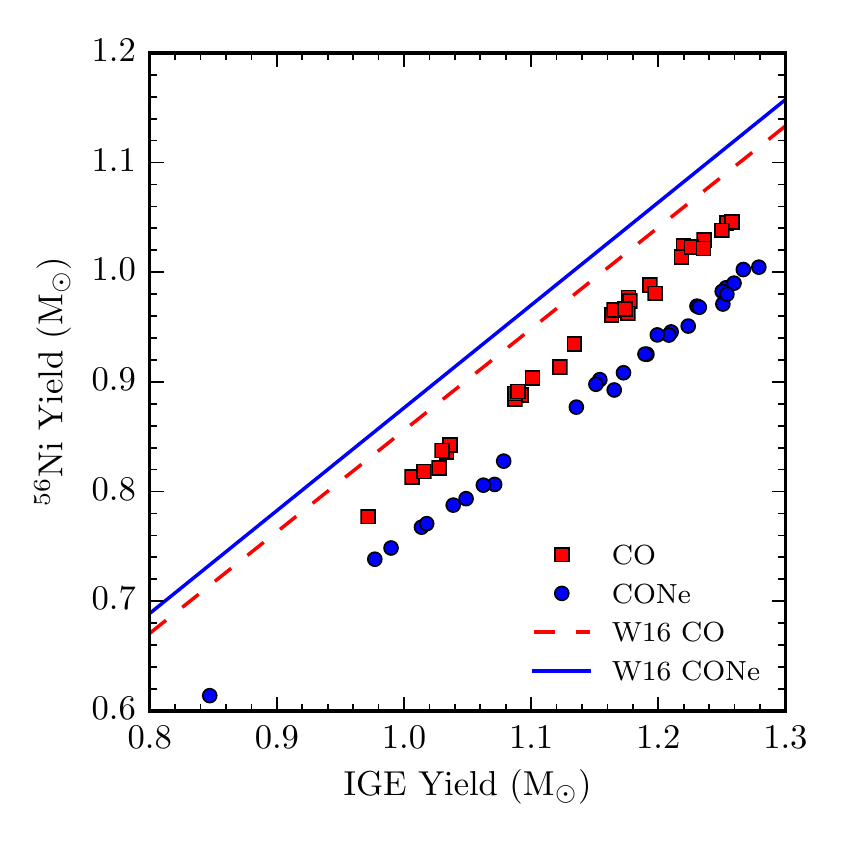}
\caption{\label{fig:conversion}
Estimated yield of \Ni{56} as a function of final mass burned to IGE
for C-O-Ne progenitors (blue) and C-O progenitors (red) at
$t = 4.0 \second$.  Lines indicate corresponding relations found in
\cite{willcoxetal2016} with
a lower central density and C-O-Ne progenitor whose interior is not mixed
during cooling.
}
\end{figure}
Figure \ref{fig:conversion} shows the estimated \Ni{56} yields
across the range of masses burned to IGE
for all C-O and C-O-Ne realizations.
The difference between the two sets of results is more pronounced
than the difference between C-O-Ne and C-O models found by \citet{willcoxetal2016}.
The results found here, in which the carbon-poor compositions have a lower fractional \Ni{56} yield, are more consistent with the expectation for a lower peak burning temperature with lower C abundance.
This motivates a closer inspection of how the \Ni{56}/IGE ratio evolves during the explosion.

\begin{figure}
\includegraphics[width=\columnwidth]{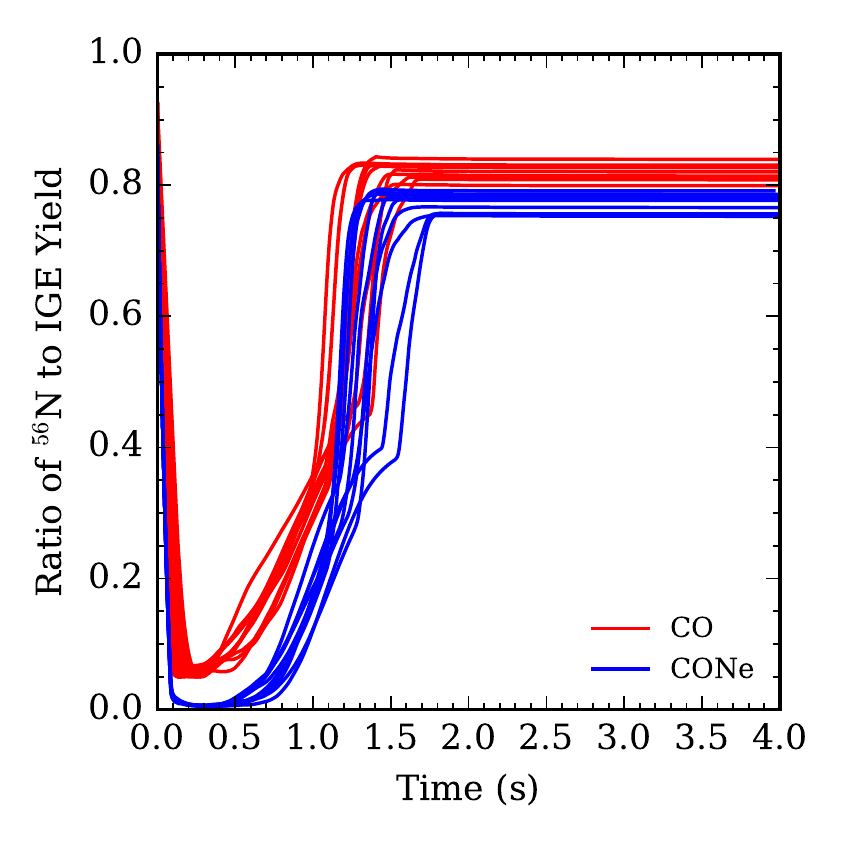}
\caption{\label{fig:compare_ratio}
Expected ratio of \Ni{56} to total IGE production by mass vs.\ time showing the evolution
for both suites of simulations. 
Shown are realizations 21-30.}
Results from C-O-Ne progenitors are in blue and C-O progenitors in red.
%SAME AS DON'S FIGURE 19
\end{figure}
\begin{figure}
\includegraphics[width=\columnwidth]{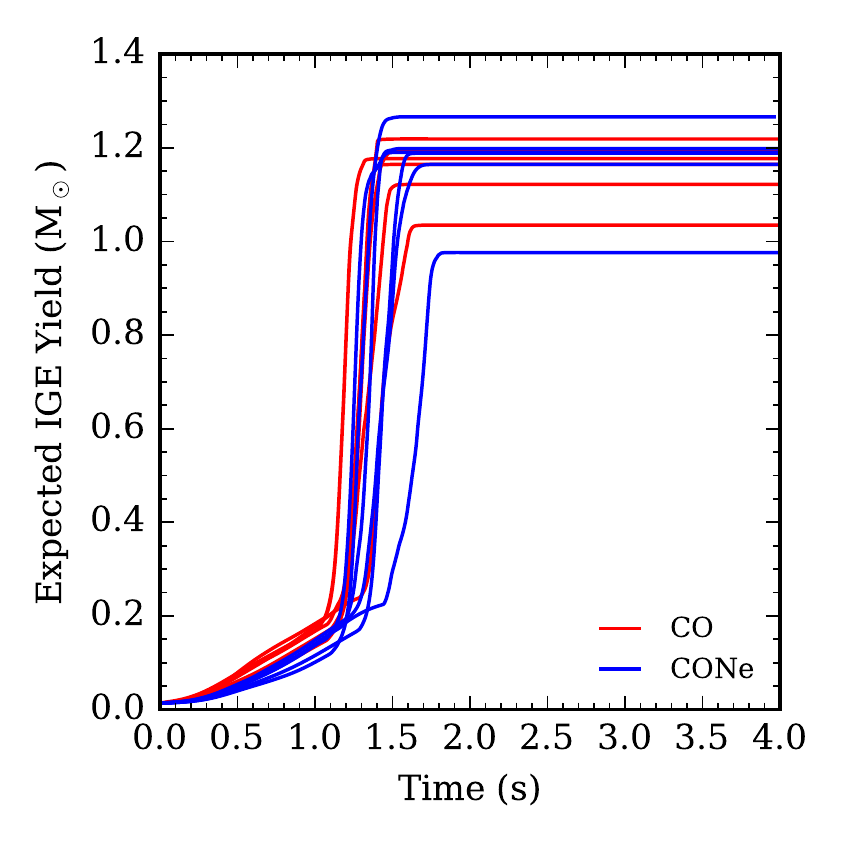}
\caption{\label{fig:compare_burned}
Plot of estimated mass burned to IGEs vs.\ time showing the evolution
for both suites of simulations for realizations 25-30. Results from C-O-Ne progenitors are
in blue and C-O progenitors red.
%SAME AS DON'S FIGURE 20
}
\end{figure}

The evolution of the fraction, by mass, of IGE material expected to be in the form of \Ni{56} after 
freeze-out is shown in Figure \ref{fig:compare_ratio}. The figure presents the results from the
same realizations as in Figure \ref{fig:nithybrid}. The ratios for the yields of 
explosions from hybrid progenitors are consistently lower than the ratios of yields from
traditional progenitors.
Figure~\ref{fig:compare_burned} shows the 
concurrent evolution of mass burned to IGE for realizations 25-30 from both suites.
The total IGE has a mostly similar evolution, and the hybrid results show
more scatter, but there is not an obvious difference between the results from
the two progenitors. These plots together show that the 
ratio is the quantity that demonstrates important differences between 
C-O and C-O-Ne cases.

During the deflagration phase, the C-O models
on average burn slightly more material to IGE and also had a significantly higher
fraction of IGE material that is \Ni{56}, yielding more \Ni{56} than
the C-O-Ne models. This result is counter to that observed in
\citet{willcoxetal2016}.
In that work, with a lower density progenitor with a deflagration ignited off-center, the C-O mostly showed a smaller fraction of IGE in the form of \Ni{56} through most of the deflagration phase.
That difference was then closed during the detonation phase, to obtain similar ratios as seen in the lines in Figure~\ref{fig:conversion} above.
A similar effect happens here during the detonation phase, but the fraction of material that is \Ni{56} stays higher for the C-O progenitors.

The differences seen in the deflagration phase in Figure~\ref{fig:compare_ratio} appear to be critical.
Central ignition at a higher central density leads to much more electron capture during the deflagration phase for the C-O-Ne case.
Remarkably, up to around 0.5 seconds, no \Ni{56} is expected to be produced by the C-O-Ne progenitor.
The comparison to the results of \citep{willcoxetal2016} indicate that the central ignition, in higher density regions, is important to this outcome.
Comparing to the C-O progenitor here indicates that the lower energy release may also slow the rise of plumes, allowing more electron capture.

\section{Discussion and Conclusion}

As mentioned above, we describe the results principally in term of the
yield of the \Ni{56}, a proxy for brightness.  Our first conclusion
is that hybrid progenitors that have experienced mixing during accretion are 
viable progenitors for type Ia supernovae. \Ni{56} yields from simulations
in DDT explosion paradigm are consistent with yields inferred from
observations, roughly 0.6 \Msun.  There is of course considerable scatter
in the results and as with other models in the DDT paradigm,
these models produce more than 0.6 \Msun, the average being 0.89 \Msun,
indicating these would correspond to very bright events.  

Going beyond the first-order, our simulations of explosions from
these hybrid progenitors gave a lower \Ni{56} yield than from
similar traditional C-O progenitors, a result expected from
the earlier study of explosions from unmixed hybrid 
progenitors~\citep{willcoxetal2016}. We also attribute this 
difference to the lower C abundance in the hybrid progenitor than
in the traditional progenitor. 

Comparing the relative abundances of \Ni{56} to IGE between the
mixed hybrid progenitors and traditional C-O models indicated
substantial differences, with the hybrid progenitors producing
a lower ratio of \Ni{56} to IGEs. This difference was particularly
pronounced during the deflagration phase of the explosion, and 
from that we conclude that the evolution of high-density material
and the amount of neutronization during the burning at high densities
is critical to the evolution and outcome of the explosion.

We were also able to compare the results from these mixed
hybrid progenitors to those of the earlier unmixed progenitors.
\citet{willcoxetal2016} reported that compared to the unmixed
hybrid progenitors, C-O models consistently yielded a greater IGE mass. 
They interpreted this result as following from 
the lower \C{12} abundance in the hybrid models and the fact that \Ne{20}-rich fuel burns to cooler 
temperatures than the fuel in traditional C-O models, which. results in slower burning and thus a lower
overall IGE yield. Our yields from explosions from the mixed progenitors did not
show the same diminution in IGE yield. We attribute this difference to the higher
central density required for ignition of these mixed hybrid models.
The results of explosions from the mixed hybrid models 
produced a smaller fraction of IGEs in the form of \Ni{56} than the unmixed hybrid case, 
and we attribute this difference 
to both the central density and the location, center or off-center, of deflagration ignition,
which determine the amount of electron capture during burning and hence the relative abundance
of \Ni{56} to IGEs.

To summarize our conclusions, we attribute the following major differences to the progenitor treated here, 
which is mixed during cooling and centrally ignited, compared to the case in \citet{willcoxetal2016} 
that did not mix while cooling:

\begin{itemize}

\item
Central ignition enhances the production of IGE.

\item
The higher central density necessitated for ignition by the lower central carbon fraction leads to a higher production of IGE as well as stronger electron capture.

\item
The off-center ignition used for the C-O-Ne progenitor in \citet{willcoxetal2016} appears to have suppressed electron capture due to the deflagration being more isolated to lower densities.

\item
In these mixed hybrid models, electron capture is noticeably enhanced in the C-O-Ne case compared to C-O at the same density and ignition location.

\item
These features together indicate that explosions from C-O-Ne progenitors should have similar IGE yields to normal C-O progenitors but lower kinetic energies.
At the same time, C-O-Ne explosions should show lower \Ni{56} yield due to enhanced electron capture during the deflagration.

\end{itemize}

The ``punch line'' of all this is that hybrid progenitors should produce dimmer
explosions, but not as much difference as found in \citet{willcoxetal2016}.
We end by noting that the existence of these hybrid
WDs is a recent state-of-the-art result from stellar evolution, but our understanding 
and ability to model stellar evolution are far from complete. These results depend
on the details of the convection and those details depend in turn on the
presence of the convective Urca process, which is still an outstanding problem 
subject to ongoing study \citep{calderetal2019,willcoxetal2019}.
The choice of central or off-center ignition is tied to how convective Urca affects the convection before the explosion.

%\software{Flash \citep{Fryxetal00,calder.curtis.ea:high-performance,calder.fryxell.ea:on,flash_pragmatic,flash_evolution} (\url{http://flash.uchicago.edu/}), 
%MESA \citep{mesa1,mesa2,mesa3,mesa3e} (\url{http://mesa.sourceforge.net/}),
%The Helmholtz EOS table used in \CASTRO\ is available in the public BoxLib
%Microphysics repository at
%  (\url{https://github.com/BoxLib-Codes/Microphysics.git}, commit
%  hash $\mathrm{45ed859b6c1dc80d831d93f9728986d6ad6e1ddc}$),
%  Matplotlib (\url{http://dx.doi.org/10.5281/zenodo.44579}).
%}

\software{Flash \citep{Fryxetal00,calder.curtis.ea:high-performance,calder.fryxell.ea:on,flash_pragmatic,flash_evolution} (\url{http://flash.uchicago.edu/}), 
MESA \citep{mesa1,mesa2,mesa3,mesa3e} (\url{http://mesa.sourceforge.net/}),
The Helmholtz EOS table used in \CASTRO\ \citep{timmes.swesty:accuracy}, Available in the public BoxLib
Microphysics repository at
  (\url{https://github.com/BoxLib-Codes/Microphysics.git}, commit
  hash $\mathrm{45ed859b6c1dc80d831d93f9728986d6ad6e1ddc}$),
  Matplotlib (\url{http://dx.doi.org/10.5281/zenodo.44579}).
}

\acknowledgements

This work was supported in part by the U.S.\ Department of Energy under
grant DE-FG02-87ER40317 and in part by the U.S.\ National Science Foundation
via a supplement to grant AST-1211563.
Support also came from the Data + Computing = Discovery summer program at
the Institute for Advanced Computational Science at Stony Brook University.
The software used in this work was developed in part by the DOE-supported
ASC/Alliances Center for Astrophysical Thermonuclear Flashes at the
University of Chicago. The authors gratefully thank Josiah 
Schwab for discussion of the study and comments on the manuscript.
thank Peter Hoeflich for very helpful discussions of this and related work. 
The authors also thank the anonymous referee for insightful commentary
that improved this manuscript.
The source code used for
this study is available as a package compatible with the
current \FLASH\ code from \url{http://astronomy.ua.edu/townsley/code.}
Results in this
paper were obtained using the high-performance computing system at the
Institute for Advanced Computational Science at Stony Brook
University.

%%%%%%%%%%%%%%%%%%%%%%%%%%%%%%%%%%%%%%%%%%%%%%%%%%%%%%%%%%%%%%%%%%
% SOFTWARE
%% \software{
%%   FLASH \citep{Fryxetal00},
%%   CASTRO \citep{castro1},
%%   MESA \citep{mesa1},
%%   Matplotlib \citep{http://dx.doi.org/10.5281/zenodo.44579}
%% }

%%%%%%%%%%%%%%%%%%%%%%%%%%%%%%%%%%%%%%%%%%%%%%%%%%%%%%%%%%%%%%%%%%
%\bibliography{master}

%ms.bbl file here for non-bibtex submit

\end{document}